\documentclass[aps,pre,reprint, amsmath, amssymb,superscriptaddress]{revtex4-1}

\usepackage{morefloats}
\usepackage{bm}
\newcommand{\beq}{\begin{equation}}
\newcommand{\eeq}{\end{equation}}

\usepackage[retainorgcmds]{IEEEtrantools}
\usepackage{graphicx,tikz,placeins}
\usepackage{mathrsfs}
\usepackage{amsmath,amssymb,amsfonts,physics}
\usepackage{color}
\usepackage{float}
\usepackage{times,txfonts}
\usepackage{nicefrac}
\usepackage[colorlinks=true,linkcolor=blue,urlcolor=black,citecolor=blue,pdfusetitle]{hyperref}
\usepackage{physics}
\usepackage{soul}
\definecolor{fe}{HTML}{812fab}

\begin{document}
\title{Nonequilibrium thermodynamics of the majority vote model }
\author{Felipe Hawthorne}
\author{M\'ario J. de Oliveira}
\affiliation{Universidade de São Paulo, Instituto de Física, Rua do Matão, 1371, 05508-090 São Paulo, SP, Brazil}
\author{Pedro E. Harunari}
\affiliation{Complex Systems and Statistical Mechanics, Department of Physics and Materials Science, University of Luxembourg, L-1511 Luxembourg City, Luxembourg}
\author{Carlos E. Fiore}
\affiliation{Universidade de São Paulo, Instituto de Física, Rua do Matão, 1371, 05508-090 São Paulo, SP, Brazil}
\date{\today}

\begin{abstract} 
The majority vote model is one of the simplest opinion systems yielding distinct phase transitions and has garnered significant interest in recent years.  However, its original formulation is not, in general, thermodynamically consistent, precluding the achievement of  quantities such as power and heat, as well as their behaviors at phase transition regimes. Here, we circumvent this limitation by introducing the idea of a distinct heat bath per local configuration, in such a way that each neighborhood value is associated with a distinct  and well-defined thermal bath. Thermodynamic properties are derived for a generic  majority vote model, irrespective of its neighborhood and lattice topology. The behavior of energy/heat fluxes at phase transitions, whether continuous or discontinuous, in regular and complex topologies, is investigated in detail. Unraveling the contribution of each local configuration explains the nature of the phase diagram and reveals how dissipation arises from the dynamics.

\end{abstract}

\maketitle

\section{Introduction}
Opinion dynamics is a crucial issue in sociophysics, encompassing several
topics, such as complex social processes, populational dynamics, decision making, elections and spreading of fake news/rumors and others \cite{castellano2009statistical}. In recent years,
distinct approaches have been proposed and investigated, aimed at tackling the key aspects
of opinion dynamics. Several of them  deal with systems presenting phase transitions, marking the existence of two regimes, one of which has a prevailing given opinion.

Among the distinct opinion systems, the majority vote (MV) model is  highlighted by its
simplicity and for exhibiting universal features of nonequilibrium phase
transitions. Its interaction mechanism is comprised of the agent's tendency to align (follow) its opinion based on the majority opinion of its nearest neighborhood \cite{de1992isotropic,pereira2005majority,PhysRevE.91.022816}.
Subsequently, generalizations of the MV model have aroused interest,
taking into account  the influence  of network topology \cite{pereira2005majority,PhysRevE.91.022816}, 
the inclusion of  distinct noises \cite{vieira2016phase,encinas2019majority},  
more states per agents \cite{brunstein99,PhysRevE.95.042304} and more recently,
inertial effects \cite{PhysRevE.95.042304,harunari2017partial,encinas2018fundamental}. 
In the latter, the presence of inertial terms have revealed a robust mechanism, affecting the classification of phase transitions even for distinct lattice topologies  \cite{encinas2018fundamental} and in systems subject to temporal disorder \cite{jesus21}. More recently, the main properties of the MV have been extensively studied in terms of entropy production signatures \cite{crochik2005entropy,noa2019entropy}.

 Stochastic thermodynamics \cite{PhysRevE.82.021120,
esposito2012stochastic,seifert2012stochastic,tome2015stochastic}
has become one of the most important topics in the realm of  nonequilibrium statistical
mechanics and an appropriate framework for describing the thermodynamic properties 
of Markovian nonequilibrium systems, having as a starting point a suitable 
 definition of production of entropy which is able to discern equilibrium
 from nonequilibrium systems \cite{schnakenberg}.  Despite previous works investigating the main properties of the MV through the entropy production \cite{crochik2005entropy,noa2019entropy},  a fundamental point of  the MV and allied voter models comprises a consistent thermodynamic  formulation  able to
 link the entropy production with the fundamental concepts of heat (heat flux) and temperature.
Aimed at overcoming such drawback, recently, a thermodynamic description for
opinion models has been proposed \cite{tome2022stochastic}, in which the idea of a distinct heat bath per neighborhood opinion configuration was introduced. Such a framework not only allows associating the dynamics to well-defined temperatures but also reconciles
the relationship between entropy production and heat flux.

In this paper, we advance on the aforementioned idea, by thoroughly investigating the thermodynamics of the majority vote model. More concretely, a general and unambiguous temperature definition
is derived, providing a way to properly investigate the behavior of  entropy production and heat fluxes
in distinct phases as well as at 
continuous and discontinuous transition regimes. The investigation is also aimed at understanding the roles of inertia and distinct topologies (regular and complex).
Given that the number of reservoirs and heat fluxes increases with the connectivity, the analysis of their roles and which of them is more representative of entropy production will be addressed and, finally, used to probe its traits across phase transitions.

This paper is organized as follows: Sec. II introduces the model and its main properties, model thermodynamics will be presented in Sec. III, and conclusions are drawn in Sec. IV.

\section{Majority vote Model and phase transition behavior}\label{sec2}
    In this section, we present an overview of  the majority vote model
    and its phase transition aspects. It consists of a simple system
    with $Z_2$ ``up-down" symmetry, in which
each microscopic configuration $\eta$ is set by the collection of
$N$ individuals $\eta\equiv (\eta_1,\eta_2,...,\eta_i,...,\eta_N)$, with
$\eta_i$ being the spin variable of site $i$ which takes the values $\pm 1$ according
to whether the spin is ``up" or ``down". With probability $1-f$,
the spin $\eta_i$ tends  to align itself with its local neighborhood majority. Conversely, with complementary probability $f$, the majority rule is not
followed.  The inertial version differs from the original one by the inclusion of a term
proportional to the local spin competing with the neighborhood.
The model dynamics are  governed by the following master
equation
\beq
\frac{d}{dt}P(\eta,t) = \sum_{i=1}^N \{w_i(\eta^i)P(\eta^i,t)
- w_i(\eta)P(\eta,t)\},
\label{eq3}
\eeq
where  $w_i(\eta)$ comprises the transition rate at which each site $i$ changes its opinion  from $\eta_i$ to $-\eta_i$, given by
\begin{equation}
  w_i(\eta)=\frac{1}{2}\left\{1-(1-2f)\eta_{i}
  {\rm sgn}(X)\right\},
\label{eq2}
\end{equation}
where $X=(1-\theta)\ell+k\theta \eta_i$, $k$ is the connectivity of a site, $\theta$ is the inertia strength, and ${\rm sgn}(X)$ is the sign function. The term \(\ell\) plays a key role in the following results, it is defined as the sum of a site's neighboring spins, $\ell \coloneqq \sum_{j}\eta_j$, and we omit the dependence on \(\eta\) for convenience. At a given configuration, all sites with \(\ell\) will become thermodynamically equivalent, defining a thermal bath. The system presents two ferromagnetic phases for small $f$. Upon raising $f$, the system yields an order-disorder phase transition, where the value of the critical point is dependent on the lattice topology and neighborhood \cite{de1992isotropic,pereira2005majority}. 
 While   phase transitions are always continuous
 for the original model ($\theta=0$) \cite{PhysRevE.91.022816},
 the inclusion of inertia can shift
the phase transition from continuous  to discontinuous depending on the lattice topology and the neighborhood \cite{PhysRevE.95.042304,harunari2017partial,encinas2018fundamental}.

Since Eq. (\ref{eq2}) states that the transition rate depends  on the sign of $X$, the flip probability (whether $1-f$ or $f$) will 
depend on the interplay between  the number of nearest neighbors $k$ and $\theta$.
For  example, for $\eta_i=-1$,  the argument of ${\rm sgn(X)}$ reads $\ell-\theta(k+\ell)$
implying that the transition $-1\rightarrow +1$, due to a neighborhood with \(\ell/(k+\ell) > \theta\), occurs with probability $1-f$ (similar to the inertialess case), whereas when \(\ell/(k+\ell) < \theta\) the probability is \(f\). Thus, we define \(\ell^* \coloneqq -k \theta \eta_i/(1-\theta)\) as the threshold value splitting the neighborhoods: All values \(\abs{\ell} > \abs{\ell^*}\) yield a transition rate equals to the inertialess case, while \(\abs{\ell} < \abs{\ell^*}\) yield \(w_i (\eta) = f\) regardless of \(\eta_i\). For completeness, the transition rate is \(1/2\) when both values are the same.

For fixed $k$, as in the present case, the phase diagram \(\theta\) versus \(f\) will be characterized by plateaus. If \(\theta\) is increased, the plateaus emerge when \(\ell^*\) has an even integer value, since it marks a regime where one additional neighborhood type \(\ell\) shifts its contribution \(f \leftrightarrow 1-f\). The plateaus can be obtained by relation
\begin{equation}
    \theta^* = \frac{2m}{k+2m}, \quad m \in \mathbb{N}.
    \label{plateau}
\end{equation}
For instance, when the connectivity is \(k = 20\), these values are
\begin{equation}
    \theta^* = \left\lbrace \frac{1}{11}, \frac{1}{6}, \frac{3}{13}, \frac{2}{7}, \frac{1}{3}, \frac{3}{8}, \frac{7}{17}, \frac{4}{9}, \frac{9}{19}, \frac{1}{2} \right\rbrace,
\end{equation}
which are later verified in the phase diagrams obtained by simulations in Figs.~\ref{fig1} and \ref{fig2}. They are the same for both regular and random-regular topologies, although the classification of the phase transition is also influenced by the topology, demonstrating that the mechanism behind the appearance of such plateaus is related to sharp shifts of contribution of each neighborhood \(\ell\).

\subsection{Entropy production}
\begin{figure}
    \centering
    \includegraphics[width=0.9\columnwidth]{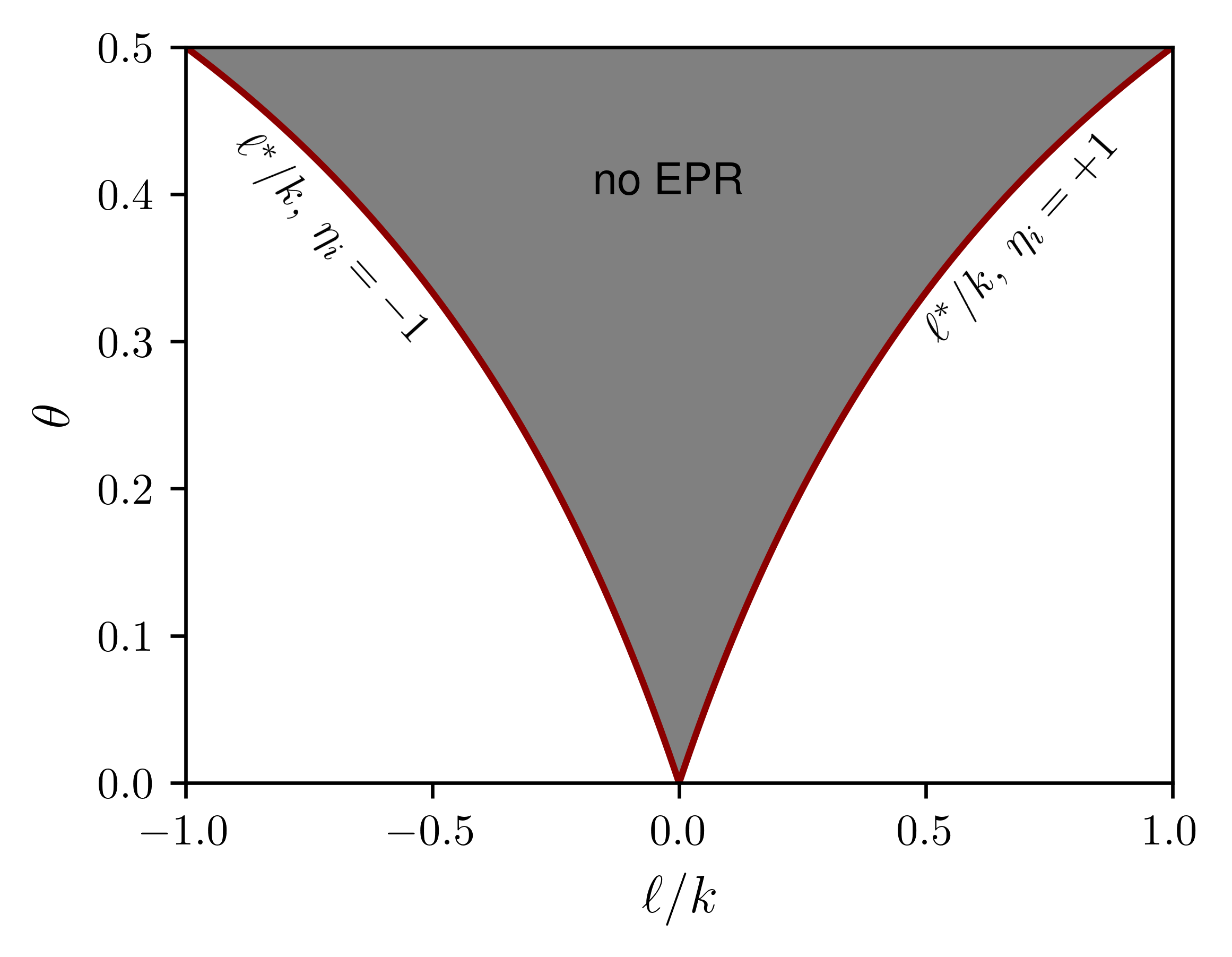}
    \caption{Scheme representing the values of \(\ell^*\) for \(\eta_i = \pm 1\) corresponding
    to the plateaus. In the shaded area, where \(\abs{\ell} < \abs{\ell^*}\), the neighborhoods do not contribute to the entropy production. }
    \label{fig:lstar}
\end{figure}

The entropy production and its connection with the heat flux is the central
issue  of this paper. Before relating both of them,  we first review the main features about the microscopic entropy production formula. Starting with the entropy definition $S=-\langle \ln P(\eta)\rangle$ (here and hereafter, we adopt the convention \(k_\text{B} = 1\) for the Boltzmann constant) and assuming the system in contact with a (or multiple) reservoir(s), its time derivative $dS/dt$ is the difference between two terms:
$dS/dt=\Pi-\sigma$, where $\Pi$ and $\sigma$ are the entropy
production and entropy flux rates,  given by the generic expressions:
\begin{equation}
\Pi = \frac12\sum_\eta \sum_i \{w_{i}(\eta^i)
P(\eta^i,t) - w_{i}(\eta)P(\eta,t)\}\ln\frac{w_{i}(\eta^i)P(\eta^i,t)}
{w_{i}(\eta)P(\eta,t)}
\label{52}
\end{equation}
and
\begin{equation}
\sigma = \frac12\sum_\eta \sum_i
\{w_{i}(\eta^i)P(\eta^i,t)-w_{i}(\eta)P(\eta,t)\} \ln\frac{w_{i}(\eta^i)}
{w_{i}(\eta)}.
\label{53}
\end{equation}
where the  one-site dynamics assumption was used. Since $dS/dt=0$
in the nonequilibrium steady-state (NESS), in which $P(\eta,t)\rightarrow p^{\rm st}(\eta)$, the steady entropy
production can be calculated from $\sigma$, acquiring the convenient  ensemble average form \cite{noa2019entropy}:
\begin{equation}
  \sigma=\sum_{i}\left\langle w_i(\eta) \ln \frac{w_i (\eta)}{w_i (\eta^{i})}\right\rangle,
  \label{eqep3}
\end{equation}
In order to evaluate $\sigma$ from Eq. (\ref{eqep3}) we take the ratio between $w_i (\eta)$
and its reverse $w_i (\eta^i)$ given by
\begin{equation}
\frac{w_i (\eta)}{w_i (\eta^i)}=
\frac{1-(1-2f)\eta_i \mathrm{sgn}[(1-\theta)\ell+k\theta\eta_i]}{1+(1-2f)\eta_i \mathrm{sgn}[(1-\theta)\ell-k\theta\eta_i]}.
\label{ep1}
\end{equation}
Inspection of the ratio above reveals that only local configurations where $\abs{\ell} > \abs{ \ell^*}$ will contribute to the entropy production. When $\abs{\ell} < \abs{ \ell^*}$, the ratio vanishes and therefore these configurations yield reversible transitions. This property is illustrated in Fig.~\ref{fig:lstar}, where values of \(\ell^*\) are shown in terms of \(\theta\). For a given finite \(k\), the even values of \(\ell\) locate the plateaus in the figure.
For $\ell \neq k\theta/(1-\theta)$, Eq.~\eqref{ep1} is conveniently rewritten as 
\begin{equation}
\ln\frac{w_i (\eta)}{w_i (\eta^i)}=-\eta_i \mathrm{sgn} (\ell)H\left[\abs{\ell}-\frac{k\theta}{1-\theta}\right]\ln (\frac{1-f}{f}),
\label{iner}
\end{equation} 
where $H(\bullet)$ is the Heaviside function. However, for $\ell=k\theta/(1-\theta)$, marking
the plateau position, Eq.~(\ref{iner}) acquires a distinct value given by $\ln ({w_i (\eta)/w_i (\eta^i)})=\eta_i \mathrm{sgn} (\ell)\ln(2f)$.
 
Above formulae  are
equivalent by calculating such a ratio
only over the subspace of local configurations in which
the ratio is different from $1$, that is for $\ell \ge k\theta/(1-\theta)$ \cite{noa2019entropy}. Thus, when expressed in terms
of the misalignment parameter $f$, the steady entropy production $\sigma$ is given by
\begin{align}
    \sigma=\frac{1}{2}\ln\frac{1-f}{f}\biggr\{ & (1-2f)\left\langle \mathrm{sgn}^2(\ell)H\left[|\ell|-\frac{k\theta}{1-\theta}\right]\right\rangle \nonumber\\ 
    &- \left\langle \eta_i \mathrm{sgn}(\ell) H\left[|\ell|-\frac{k\theta}{1-\theta}\right]\right\rangle\biggr\},
\label{eqw1}
\end{align}
which only depends on $f$ and on  $\langle \eta_i \mathrm{sgn} (\ell)H\left[|\ell|-k\theta/(1-\theta)\right] \rangle$ and
 $\langle \mathrm{sgn}^2(\ell)H\left[|\ell|-k\theta/(1-\theta)\right]\rangle$. 

\subsection{Overview about phase transitions and finite-size scaling}
As stated broadly in the literature,  continuous and discontinuous phase transitions  become rounded at
the vicinity of phase transitions  due to finite size effects, whether for equilibrium \cite{landau2021guide}
and nonequilibrium systems \cite{fiorefss,fiorefss2}.
Despite the order parameter
and its moments have been broadly exploited for characterizing
nonequilibrium phase transitions, recently, the behavior of entropy production and allied quantities (e.g. its first derivative)  has attracted a great deal of attention as their
 identificators  \cite{crochik2005entropy,noa2019entropy,goes20,seara2021irreversibility,martynec2020entropy,aguilera2021unifying}. 

According to the finite-size scaling (FSS) theory, 
 at the vicinity of the critical point $f_c$, a given quantity $X$
 [$X\in (|m|,\chi$ and $\sigma'\equiv d\sigma/df)$]
 will  behave as
$X=N^{y_x/\nu}f_x(N^{1/\nu}|\epsilon|)$, where $f_x$
is a scaling function, $\epsilon=(f-f_c)/f_c$ is the distance to
the criticality and $y_x$ is the critical
exponent obtained from ($y_x=-\beta,\gamma$ and $\alpha$)\cite{landau2021guide}.  The last exponent is similar to the relationship between the thermal derivative of the entropy, $S$, and specific heat, $C$, in equilibrium phase transitions (recalling that $C=N^{\alpha/\nu}f_c(N^{1/\nu}|\epsilon|$) \cite{landau2021guide}, illustrating that the connection between entropy production and exchanged heat presented here introduces a physical argument for such  scaling behavior.

Since the scaling behavior of heat fluxes (and their derivatives) at the criticality was considered previously in \cite{tome2022stochastic}
 we are going to focus on nonequilibrium discontinuous  phase transitions in this paper. For a generic ensemble average $X$, the starting point consists of assuming a bimodal Gaussian distribution, centered at $\mu_o$ and $\mu_d$ (with associated variances $\chi_o$ and $\chi_d$). In the case of  the steady entropy production at the vicinity of $\epsilon=f-f_c$, a bimodal entropy production
probability distribution centered at $\mu_o$ and $\mu_d$ (with associate variances $\chi_d$ and $\chi_o$) leads
to the approximate expression for $\sigma$:
 \begin{equation}
 \sigma \approx \frac{\mu_o+\overline{\alpha}\mu_d e^{-N[(\mu_o-\mu_d)\epsilon]}}{1+\overline{\alpha}e^{-N[(\mu_o-\mu_d)\epsilon]}},
\label{opr}
\end{equation}
where  $\overline{\alpha}=\sqrt{\chi_d/\chi_o}$.  We note that the ordered and disordered phases are favored as $\epsilon<0$ and $\epsilon>0$ (assuming that $\mu_o<\mu_d$), respectively, and $\sigma=(\mu_o+\overline{\alpha}\mu_d)/(1+\overline{\alpha})$ at $\epsilon=0$, indicating that all entropy production curves, simulated for distinct $N$'s, will  cross at 
the transition point $f_c$.
Having $\sigma$, its derivative in respect to $f$ behaves at the vicinity of $f_c$
as:
\begin{equation}
 \sigma'\approx   \frac{N (\mu_o-\mu_d)^2 e^{N (\mu_o-\mu_d)\epsilon}}{\overline{\alpha}\left(1+\overline{\alpha}e^{N (\mu_o-\mu_d)\epsilon}\right)^2},
\end{equation}
showing that $\sigma'$ scales with $N$
at the coexistence $\epsilon=0$, in agreement  with the above finite size expression for the quantity $X$.
Alternatively (and analogously), Eq. (\ref{opr}) is obtained  by resorting
to the ideas presented in \cite{hanggi1984bistable,PhysRevE.104.064123,basile}, where coexisting phases are treated via a two-state model in  which ordered and disordered phases are given by transition rates exhibiting an exponential dependence on the system size $N$ and proportionality to the distance  $\epsilon$
to the transition point:
\begin{equation}\label{ab}
    a \sim k\sqrt{\chi_a}e^{-N (c_0 - c_a \epsilon)},
    \qquad 
    b\sim k\sqrt{\chi_b}e^{- N(c_0 + c_b \epsilon)},
\end{equation}
where $k,c_0,c_a,c_b >0$ are constants.  ``Ordered"  and ``disordered" probabilities, $p$ and $q$ respectively, are related to rates $a$ and $b$
by means of relations $p=b/(a+b)$ and $q=1-p$, given by $p = \sqrt{\chi_b}(\sqrt{\chi_b}+\sqrt{\chi_a}e^{c N \epsilon})^{-1}$, where $c=c_a+c_b>0$. 
As shown in Ref. \cite{PhysRevE.104.064123},  a given ensemble average (including the  entropy production $\sigma = \langle \mathcal{\sigma}_\tau\rangle /\tau$  averaged over a sufficiently long time $t \rightarrow \infty$ and over many independent stochastic trajectories has its average given  by $\sigma=\mu_a p+\mu_b b$, where 
\begin{equation}
    \sigma \approx\frac{\mu_b\sqrt{\chi_b}+\mu_a \sqrt{\chi_a}e^{c N \epsilon}}{\sqrt{\chi_b}+\sqrt{\chi_a}e^{c N \epsilon}},
\end{equation}
which has precisely the form of Eq. (\ref{opr}). 

\begin{figure}
\hspace*{-0.6cm}     
 \includegraphics[scale=0.35]{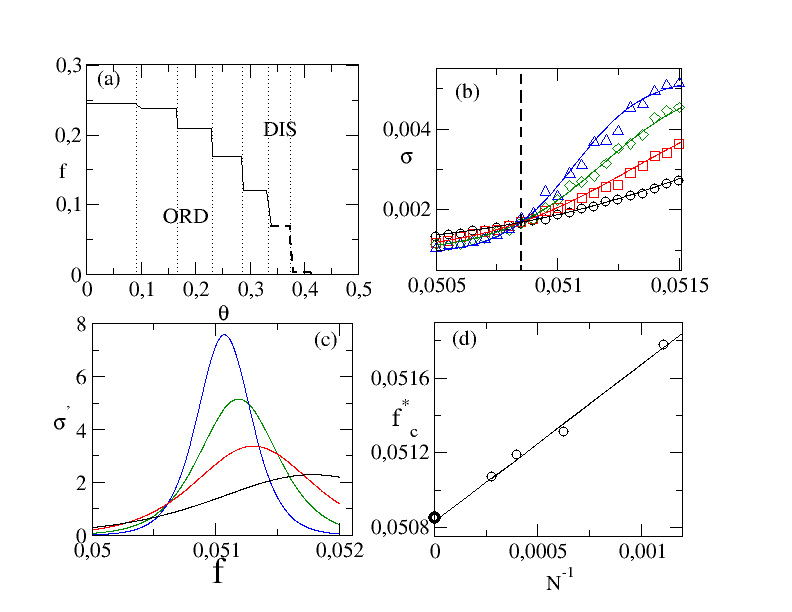} 
\caption{In $(a)$, the phase diagram of  the inertial majority model for a regular lattice for $k=20$. Vertical lines mark the plateau positions predicted in
Eq.~(\ref{plateau}). Panel $(b)$ depicts
the entropy production $\sigma$ for distinct system sizes $N=L^2$'s.
Continuous lines denote the phenomenological description
from Eq. (\ref{opr}) and vertical line corresponds to the crossing among entropy production curves at
$f_c=0.05085(2)$. In $(c)$, the derivative $\sigma'\equiv d\sigma/d f$
versus $f$ obtained from continuous lines in $(b)$. 
Panel $(d)$ show the position $f^*_c$ of maximum of $\sigma'$ versus $N^{-1}$
and  its accordance with the crossing among entropy production curves yielding (symbol $\bullet$) as $N\rightarrow \infty$.}
\label{fig1}
\end{figure}

The main features of discontinuous phase transitions are summarized in Fig. \ref{fig1}.
From now on we shall consider $k=20$
which exhibits a discontinuous phase transition for $\theta>1/3$, as depicted in panel Fig.\ref{fig1}$(a)$. Aforementioned portraits are exemplified in panels $(b)-(d)$ for $\theta=3/8$. We remark that continuous lines, given by Eq. (\ref{opr}), describe very well the behavior of the entropy production and its derivative, the latter presenting a maximum at $f^*_c$ scaling with $N^{-1}$, whose value as $N\rightarrow \infty $ agrees very well with  those obtained from the crossing among curves.

\subsection{Discontinuous phase transitions in complex topologies}
\begin{figure}
\hspace*{-0.6cm}     
\includegraphics[width=\columnwidth]{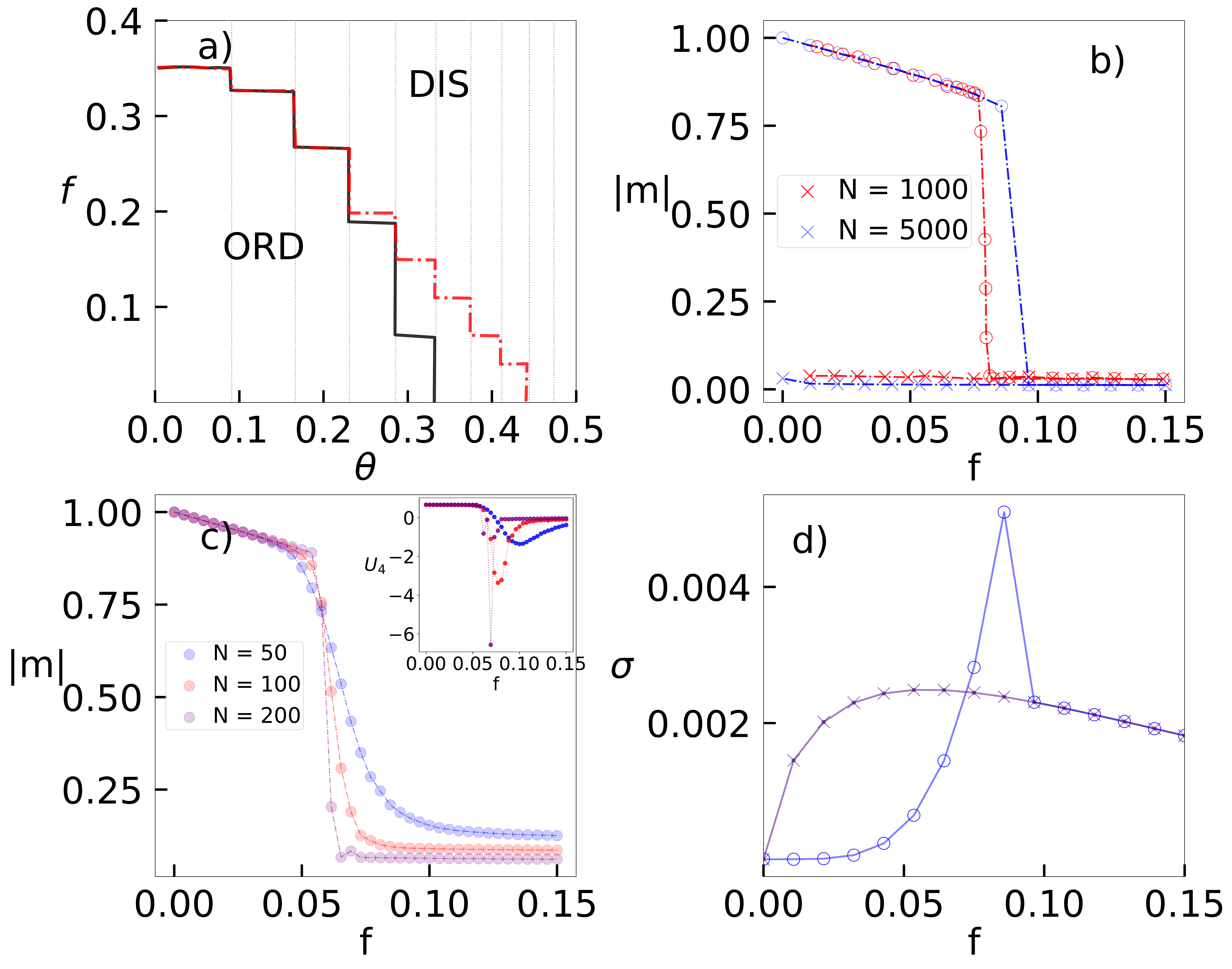} 
\caption{Phase transition for the MV in a random-regular topology with
connectivity $k=20$.  Panel $(a)$ depicts the phase diagram $\theta$ versus $f$. Continuous and dashed lines show, for a system of size $N=10^4$.  Note that a hysteretic branch for $\theta>3/13$. Panels $(b)$  and $(c)$ show, for $\theta=3/8$,  the order parameter $|m|$ versus $f$ for distinct large and small system sizes $N$, respectively. Inset:  the reduced cumulant $U_4$ versus $f$. Circles and $\times$ attempt to
the forward and backward ``trajectories", respectively. In $(d)$, its corresponding  $\sigma$'s for $N=5000$.}
\label{fig2}
\end{figure}

The behavior of discontinuous phase transitions in complex topologies is more revealing and it is different for small and large system sizes. In the former case, quantities change smoothly 
as $f$ is varied (see e.g. Fig. \ref{fig2}$(c)$), in similarity with the behavior in regular structures, also characterized by the reduced cumulant $U_4$ presenting a  minimum value increasing with $N$ (inset) and a maximum behavior of $\chi$ near
the coexistence. 
Conversely, the  behavior becomes akin to the mean-field when $N$ is large, in which the phase coexistence manifests itself by means of a hysteretic branch, e.g. a region located at 
$f_b< f< f_c$ when the dynamics evolve to the ordered (stable for $f\le f_b$) and disordered (stable for $f\ge f_c$) phases depending on the initial condition. Such changes upon raising $N$ share
some similarities with the metastable behavior observed in the dynamics and thermodynamics of work-to-work transducers, where the system behavior ``quickly" approaches the MFT's [behavior] as $N$ increases \cite{herpich}.

Here, we describe a
brief (non-rigorous) argument about the expected behavior in complex topologies by resorting
 to the ideas from Ref.\cite{noa2019entropy}. Since spins are independent of each other in the disordered phase, the  order parameter  behaves as
  $\langle\eta_i\rangle \sim N^{-1/2}$ and then a 
$n-$th correlation  will behave as
$\langle\eta_i\eta_{i+1}...\eta_{i+n}\rangle \approx \langle\eta_i\rangle \langle\eta_{i+1}\rangle...\langle \eta_{i+n}\rangle=N^{-n/2}$.  Hence,
in the thermodynamic limit,  all correlations will vanish  
 and $\sigma$ will depend solely on control parameters. On the other hand,
 $\langle\eta_i\eta_{i+1}...\eta_{i+n}\rangle$ is expected to be finite and also $f-$ dependent in the ordered phase, consistent with $\sigma$  exhibiting a dependence  on the control parameters and correlations. Therefore,
 the existence of a hysteretic loop for the order parameter (panel $(b)$) is also translated to the entropy production behavior (see e.g. panel $(d)$).
\section{Thermodynamics of the majority vote model}
\subsection{General features}
In Sec. \ref{sec2}, the main properties of the majority vote model were analyzed without any thermodynamics consideration. In this
section, we address this issue in detail. 
The central point consists of assuming that the
one-site transition rate $w_i(\eta)$ is decomposed in
$\ell$ distinct (mutually exclusive) components, each one associated with a given thermal bath 
 (reciprocal inverse temperature $\beta_\ell$),
given by $w_i(\eta) = \sum_\ell w_{\ell i}(\eta)$ ($\ell=2,4,...,k$), where $ w_{\ell i}(\eta)$ assumes the Glauber form:
\begin{equation}
    w_{\ell i}(\eta)=\frac{\alpha_\ell}{2}\{1-\tanh(\beta_\ell \Delta E) \},
    \label{eqg}
\end{equation}
where $\alpha_\ell$ is a constant and  $\Delta E=E(\eta^i)-E(\eta)$ denotes 
the energy difference between configurations $\eta$ and $\eta^i$. For ``up-down" $Z_2$ symmetry systems, the energy  can be generically expressed according to the Ising-like
form
$E(\eta)=-J\sum_{(i,j)}\eta_i\eta_j-H\sum_{i=1}^N\eta_i$ \cite{yeomans1992statistical},
where $J$ represents the interaction energy between spins, and
 $H$ is a parameter accounting for the dependence on the local spin $\eta_i$ (usually  the magnetic field). Giving that $\mathrm{sgn}(\ell)=-\mathrm{sgn}(-\ell)$, one has $H=0$ for all values of $\theta$.  From  Eq. (\ref{eqg}), 
 the ratio $ w_{\ell i}(\eta^i)/ w_{\ell i}(\eta)$
 is  then expressed according to the local detailed balance:
\begin{equation}
    \frac{w_{\ell i}(\eta)}{w_{\ell i}(\eta^i)}
=  e^{-\beta_\ell [E(\eta^i)-E(\eta)]}.
\label{23}
\end{equation}
We are now in a position to obtain the model's thermodynamic properties.
The time variation of the mean energy $U=\langle E(\eta)\rangle$  is given by  $dU/dt = \sum_\ell \Phi_\ell$, where $\Phi_\ell$ denotes the heat exchanged due
to the $\ell$-th thermal bath, given by
\begin{equation}
\Phi_\ell = 
\sum_i \langle [E(\eta^i)
- E(\eta)] w_{\ell i}(\eta)\rangle,
\label{4}
\end{equation}
constrained by the first law of thermodynamics, $\sum_\ell \Phi_\ell =0$ in the NESS. The entropy production and entropy flux are also decomposed into \(\ell\)-indexed components by replacing \(w_{i}(\eta)\rightarrow w_{\ell i}(\eta)\) in Eqs.~\eqref{52} and \eqref{53}. In particular, the latter reads
\beq
    \sigma_\ell = \sum_\eta p^{st}(\eta) \sum_i
    w_{\ell i}(\eta) \ln\frac{w_{\ell i}(\eta)}
    {w_{\ell i}(\eta^i)}.
    \label{53_2}
\eeq
Since the entropy change vanishes at the NESS, \(dS/dt = \sum_\ell(\Pi_\ell -\sigma_\ell)\), both entropy production and entropy flux can be identified by Eq.~\eqref{53_2}: $\Pi=\sum_\ell \Pi_\ell=\sum_\ell\sigma_\ell=\sigma$. The expressions above are consistent with Refs. \cite{seifert2012stochastic,van2015ensemble}.

Finally, by inserting Eq. (\ref{23}) into Eq. (\ref{53_2}), each entropy flux component $\sigma_\ell$ is related with exchanged heat $\Phi_\ell$ by a Clausius-like form
$\sigma_\ell=-\beta_\ell\Phi_\ell$, where $\Phi_\ell$  is given by Eq. (\ref{4}).
Alternatively, $\sigma$ can also be written in the usual  thermodynamics form as a  sum of thermodynamic fluxes times forces:
\begin{equation}
 \sigma=-\sum_{\ell}\beta_\ell\Phi_\ell \qquad {\rm or} \qquad \sigma=\sum_{\ell\neq 2}X_\ell \Phi_\ell,   
 \label{entrot}
\end{equation}
where the second temperature was set as a reference to define all $(k/2)-1$ thermodynamic forces $X_\ell\equiv\beta_2-\beta_\ell$, associated with its respective flux, $\Phi_\ell$.
For simplicity, we set the Ising interaction parameter to $J=1$. From the expression for $E(\eta)$, it follows that $\Delta E=2 \eta_i\ell$, which can be rewritten as $\Delta E=2 \eta_i |\ell|\mathrm{sgn}(\ell)$. By taking the logarithm of Eq. (\ref{23}), it follows that 
\begin{equation}
    \ln\frac{w_{\ell i}(\eta)}{w_{\ell i}(\eta^i)}=-2\beta_\ell |\ell|\eta_i \mathrm{sgn} (\ell).
\end{equation}
Since the transition rates associated with each thermal bath
are mutually exclusive, a direct comparison with Eq. (\ref{iner}) for a given $\ell$ provides to obtain each (reciprocal inverse) temperature $\beta_\ell$ given by
\begin{equation}
    \beta_{\ell}=\frac{1}{2|\ell|}H\left[|\ell|-\frac{k\theta}{1-\theta}\right]\ln(\frac{1-f}{f}), 
    \label{temp}
\end{equation}
where $\beta_2=2\beta_4=3\beta_6...=k\beta_{k}/2$ in the inertialess case. We pause to make a few comments. First, 
 Eq. (\ref{temp}) is one of the main results of this paper, it extends the calculation of temperatures for a given neighborhood
 and inertia, and reduces to the expression from Ref. \cite{tome2022stochastic} as
 $\theta=0$. Second, $\beta_\ell$ vanishes for large enough values of inertia $\theta>\theta_p$, illustrating that despite a heat flux associated with the $\ell$-th reservoir being well-defined, it does not produce entropy. Third and last, the temperature assumes a different value for $\theta=\theta_p$ given by
\begin{equation}
    \beta_{\ell}=-\frac{1}{2|\ell|}\ln(2f).
    \label{temp2}
\end{equation} 
This completes our  description of the temperature definitions for the MV as well as  the influence of inertia.
Now we turn to unravel the role of each \(\ell\) to the fluxes of heat and entropy production.
Starting with the inertialess case, where $\beta_2>\beta_4>...>\beta_k$, we argue that the heat fluxes associated with the states in contact with the coldest and hottest baths are always positive and negative, respectively: $\Phi_2<0$ and $\Phi_k>0$, whose a (non-rigorous) argument is present as follows.
  Starting with the two thermal
 baths case ($k=4$), it is straightforward to verify that, since  $\sigma$ acquires the simple form  $\sigma=(\beta_2-\beta_4)\Phi_4>0$. Given that $\beta_2-\beta_4>0$ [cf. Eq.~(\ref{temp})],
 it follows that $\Phi_4\ge 0$ and hence $\Phi_2=-\Phi_4\le 0$. 
  The case of more than two reservoirs is more intriguing, since 
  intermediate fluxes can be positive, negative, or even change  their sign upon $f$ being varied [see e.g. Fig. \ref{fig3} $(d)$].  
For $k=6$, one has $\sigma=-(\beta_2-\beta_6)\Phi_2-(\beta_4-\beta_6)\Phi_4\ge 0$ and three possibilities for $\Phi_2$ and 
 $\Phi_4$.  The former, in which  both are negative, promptly
 implies $\sigma\ge 0$, whereas the  second case,  $\Phi_2\le 0$ and 
 $\Phi_4\ge 0$, is also consistent since $-(\beta_2-\beta_6)\Phi_2\ge (\beta_4-\beta_6)\Phi_4$ and hence  $\Phi_6\ge 0$ (recalling that  $\Phi_6=-(\Phi_2+\Phi_4)$). The third possibility, in which  $\Phi_2\ge 0$ and 
 $\Phi_4\le 0$  violates the second law in some cases and thus it is not possible.
 Similar findings are verified for $\theta\neq 0$, but we should note that only neighborhoods with $\ell^*$ greater than   $\ell-k\theta/(1-\theta)$   will contribute to the entropy production, 
 $\sigma=-\sum_{\ell^*}^k\beta_{\ell} \Phi_{\ell}$. For example, for 
 $k=20$ and distinct inertia intervals  $3/8<\theta\le 7/17$, $7/17<\theta\le 4/9$,  $\theta>4/9$,
the corresponding entropy productions read $\sigma=-\sum_{\ell=14}^k\beta_{\ell}\Phi_\ell$, $\sigma=-\beta_{16}\Phi_{16}-\beta_{18}\Phi_{18}-\beta_{20}\Phi_{20}$ and  $\sigma=-\beta_{18}\Phi_{18}-\beta_{20}\Phi_{20}$, such latter one
 similar to the $k=4$ case (but here $\sum_{\ell=2}^k\Phi_\ell=0$) and once again illustrating that 
$\Phi_{\ell^*=18}\le 0$ and $\Phi_{k=20}\ge 0$. We close this section by pointing out that, despite the above non-rigorous argument, the general
finding $\Phi_{\ell^*}\le 0$ and $\Phi_{k}\ge 0$ has
been verified  in all cases. In contrast, it is not possible to draw general conclusions about
 intermediate fluxes, in which some change sign as $f$ increases.

\subsection{Fluctuation theorems}

Thermodynamic consistent systems satisfy the detailed fluctuation theorem (DFT) for entropy production, which gives rise to the stochastic version of the second law. It states that negative fluctuations of the integrated entropy production are exponentially suppressed by the positive counterparts. For a given integration window \(\tau\), the DFT is asymptotically valid for \(\Sigma = \int_0^\tau \sigma(t) \mathrm{d}t\) at the NESS since it is equal to the entropy production:
\begin{equation}\label{DFT}
    \lim_{\tau\to\infty} \ln \frac{P_\tau (\Sigma)}{P_\tau (-\Sigma)} = \Sigma,
\end{equation}
where \(P_\tau(\Sigma)\) represents the probability of measuring \(\Sigma\) in a trajectory of length \(\tau\). This relation holds beyond the long-time limit when the internal change of configuration entropy is considered in addition to the entropy fluxes. Consequence of the above, the integral fluctuation theorem (IFT) reads
\begin{equation}\label{IFT}
    \lim_{\tau\to\infty} \expval{e^{-\Sigma}}_\tau = 1
\end{equation}
and is useful for relating the components of \(\Sigma\), such as in the celebrated Jarzynski equality \cite{j1} that relates the statistics of work to free energy differences, bridging equilibrium and nonequilibrium quantities. The feasibility of employing such methods is tightly related to the ability to observe fluctuations in the trajectories, which become rare as \(\tau\) increases. We explore the manifestation of these relations, cornerstones of stochastic thermodynamics, in the MV vote model.

\begin{figure}
    \includegraphics[width=.5\columnwidth]{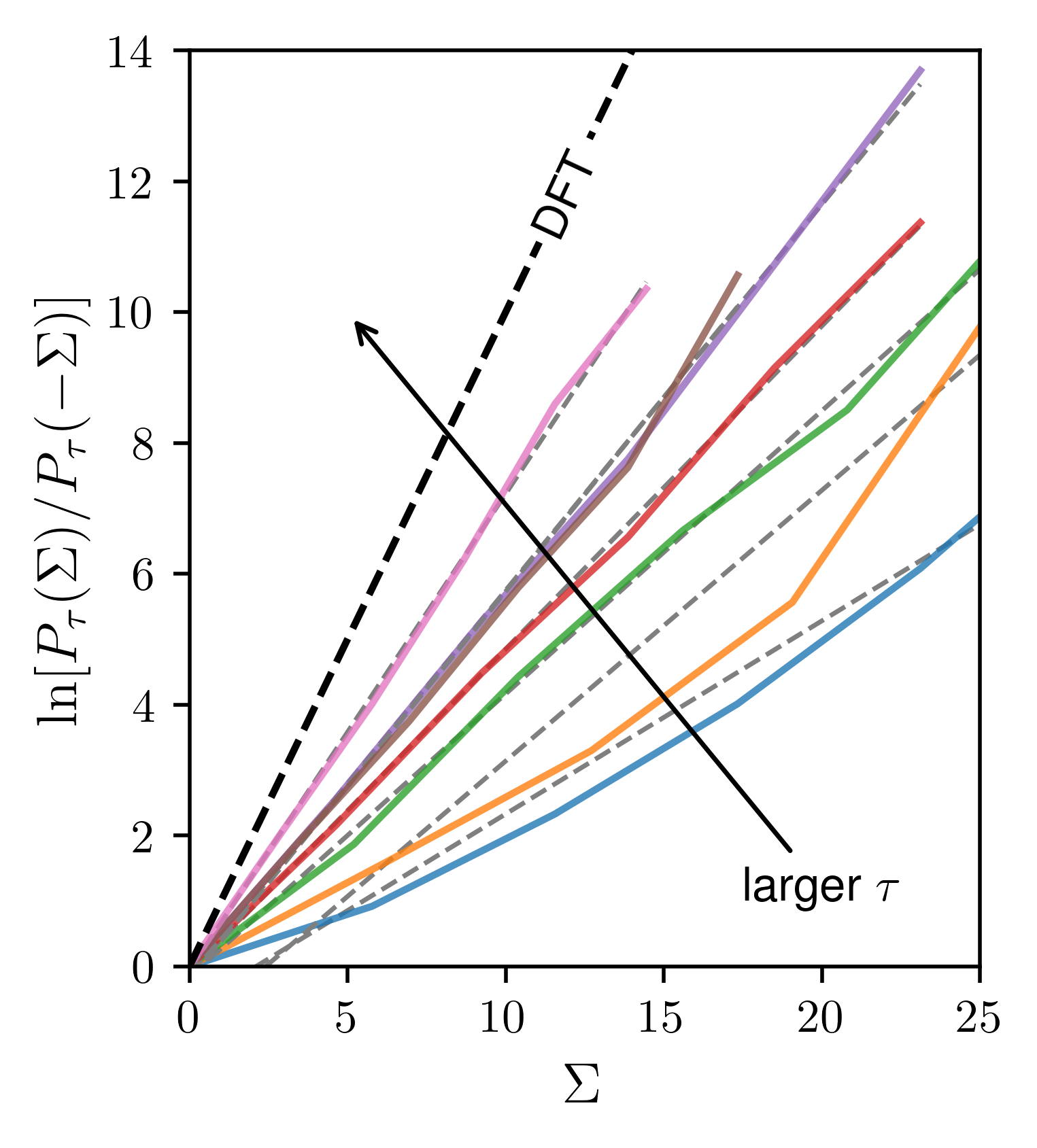}%
    \includegraphics[width=.5\columnwidth]{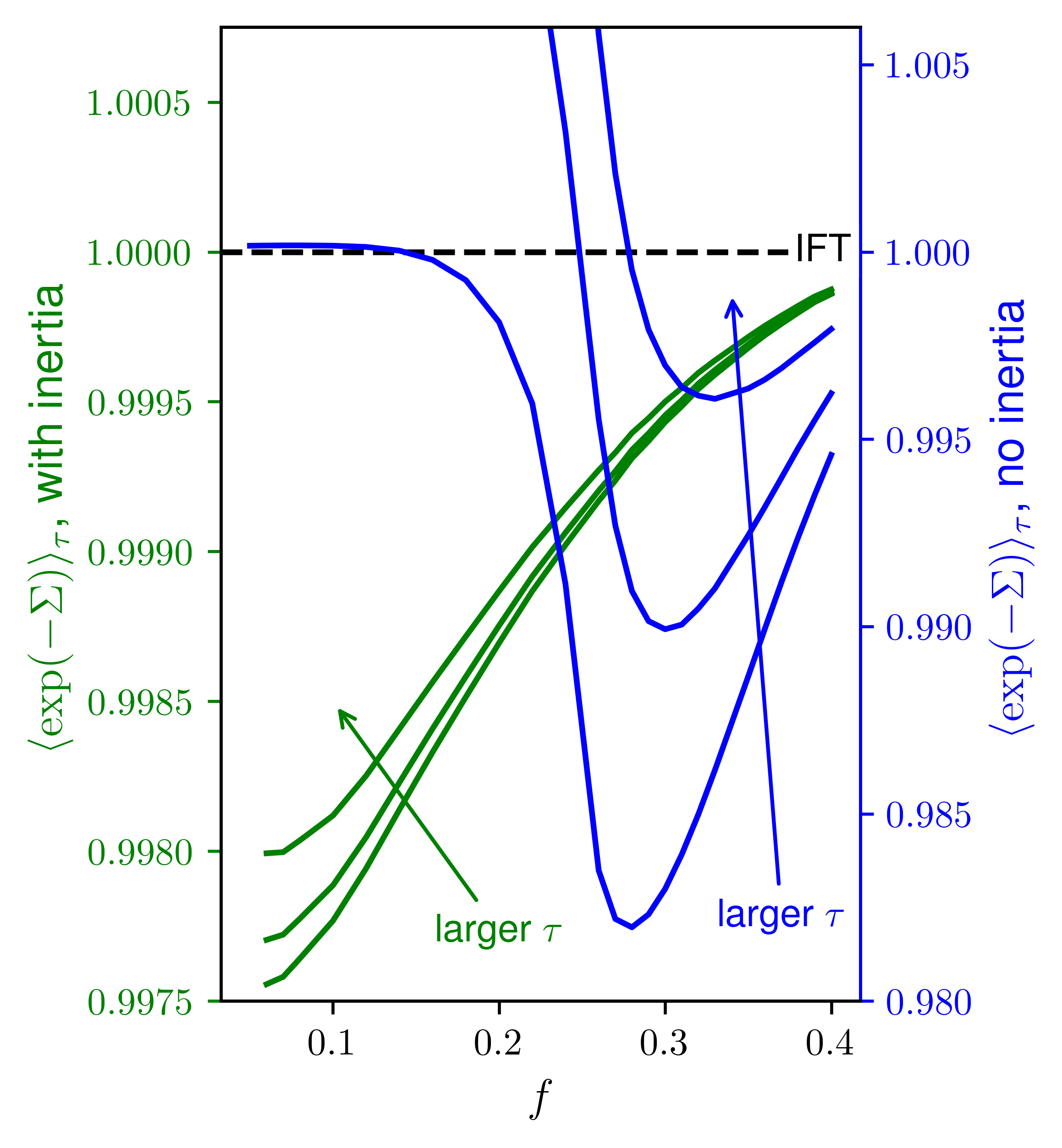} 
    \caption{ \textit{Left}: Convergence to the detailed fluctuation theorem as integration window \(\tau\) increases for a lattice \(L=6\) and \(f=0.04\); solid lines are simulation results while dashed lines are the respective linear fits. \textit{Right}: Convergence to the integral fluctuation theorem for the case with no inertia (blue) and with inertia \(\theta = 3/8\) (green); additional parameters are \(k=20\) and \(N = 10^4\).}
\label{fig0}
\end{figure}

The left panel of Fig.~\ref{fig0} shows the convergence of the left-hand side of Eq.~\eqref{DFT} to its right-hand side as the integration windows get larger for the entropy production
evaluated from Eqs. (\ref{53_2}) and (\ref{entrot}). Observing the DFT becomes an expensive task even for small systems since the negative fluctuations of entropy production become increasingly rare for larger values of \(\tau\). The right panel shows the left-hand side of the IFT in Eq.~\eqref{IFT}, which converges to one despite the presence of inertia. It is worth mentioning that the convergence is observed from above and from below. Although no general conclusion can be drawn, the behavior of these fluctuation relations might be related to the phase transitions: In the examples, the IFT presents a slower convergence at the vicinity of the phase transition.

\begin{figure}
\hspace{-0.4cm}
 \includegraphics[width=\columnwidth]{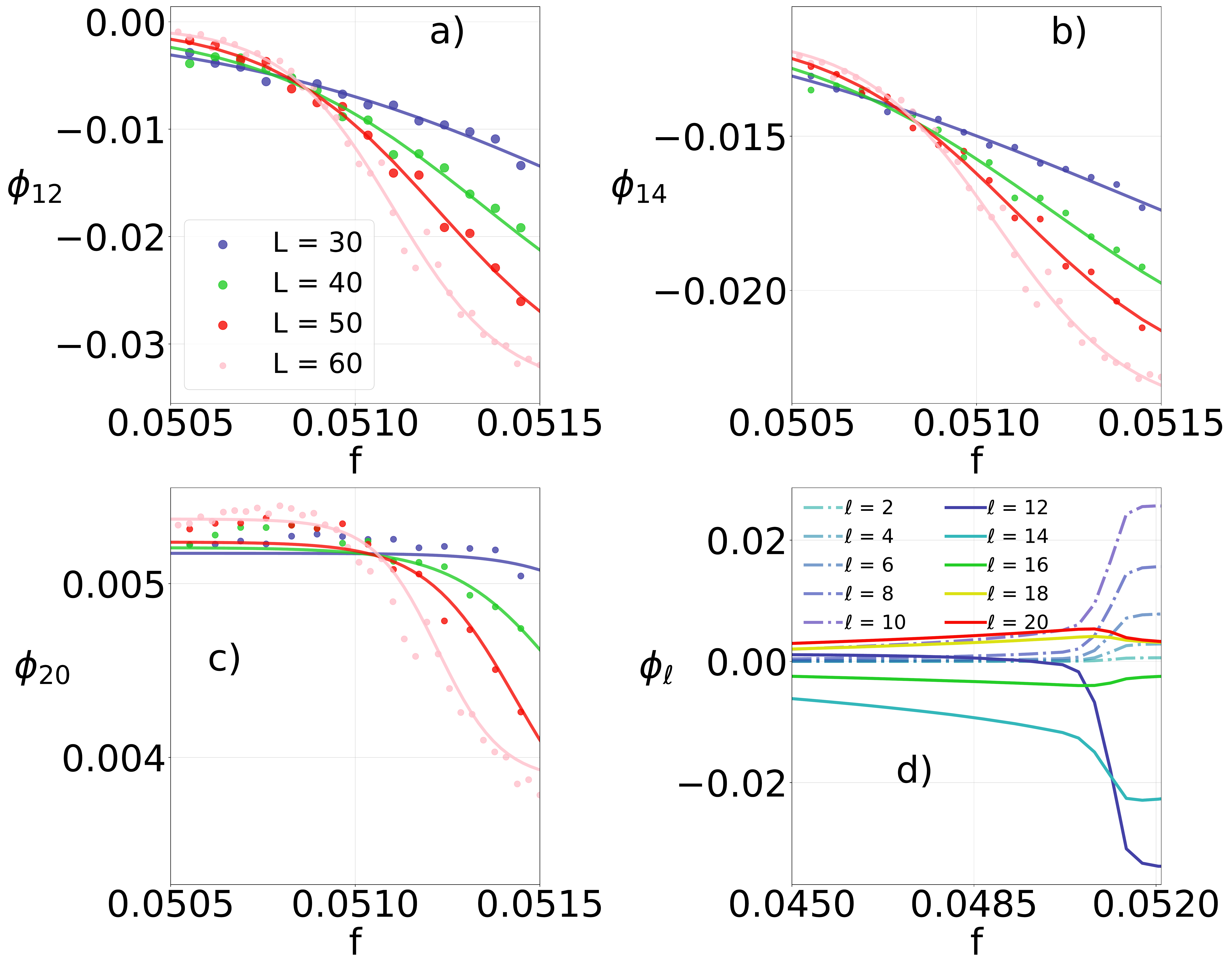} 
\caption{For the regular lattice with $\theta=3/8$,  $k=20$ and distinct system sizes $N=L^2$, panels (a)-(b)
depict the most representative (largest absolute values) heat fluxes  per particle $\Phi_\ell$'s 
versus control parameter $f$. Continuous lines denote correspond to the phenomenological
approach according to the ideas of Eq. (\ref{opr}). Although the component heat flux panel $(c)$ mildly changes with $f$, all curves also cross at $f_c$.
Panel (d) shows all $\Phi_\ell$'s ($\ell=2,4,...,k$) for $N=60^2$.  }
\label{fig3}
\end{figure}

\begin{figure}
\hspace*{-0.4cm}     
 \includegraphics[width=\columnwidth]{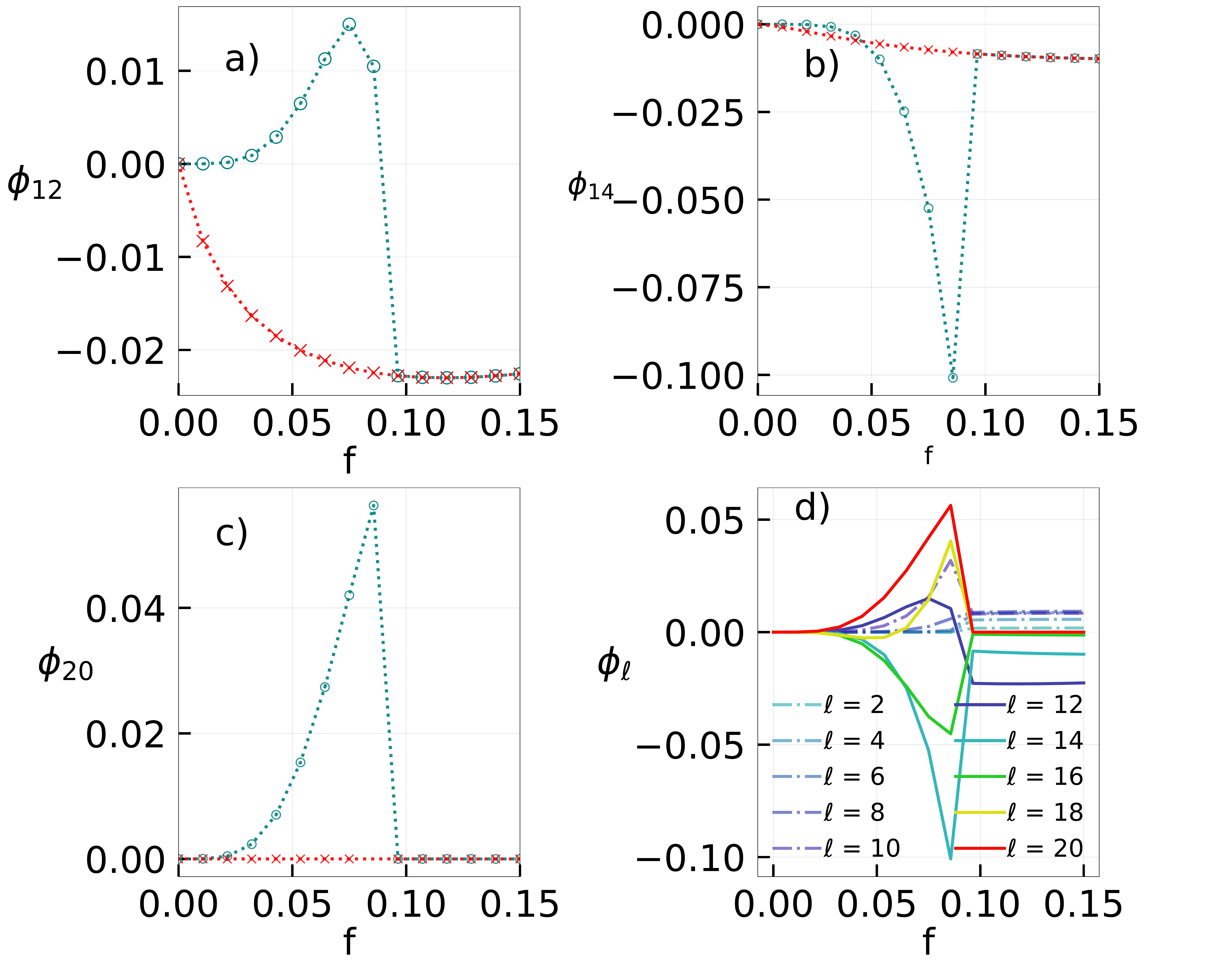} 
\caption{For a system of size $N=5000$,  the same as before, but for a random-regular structure.}
\label{fig3RR}
\end{figure}

\subsection{Heat fluxes at phase transitions }
According to Eq. (\ref{4}), every heat flux $\Phi_\ell$ is an ensemble average and, therefore, we expect at least the most significant components of the entropy production to behave similarly to $\sigma$ at the vicinity of a phase transition. More specifically, at discontinuous phase transitions, the curves of entropy production cross at \(f_c\) for distinct system sizes in regular lattices, and a hysteretic branch is present in complex topologies \cite{noa2019entropy}. These properties are promptly verified for the largest fluxes \(\ell = 12\), 14 and 20. For a regular lattice, panels (a)-(c) of Fig.~\ref{fig3} display the crossing of the fluxes for different systems sizes, and panel (d) shows the quantitative value of each individual flux. For a random-regular network, panels (a)-(c) of Fig.~\ref{fig3RR} show the hysteretic branch while (d) shows individual flux values.

The continuous lines in panels Fig.~\ref{fig3} (a)-(c) are obtained from the bimodal Gaussian description in Eq.~\eqref{opr}, in good agreement with the simulation results. Remarkably, for both regular and complex topologies, the phase transition can be probed and precisely located from the behavior of any individual flux.

\subsection{Contributions to dissipation}

Inspecting the thermodynamic contribution of individual \(\ell\)'s raises the question of how each type of neighborhood contributes to entropy production, a measure of dissipation. As previously discussed, the second law imposes $\Phi_{\ell^*}< 0$ and $\Phi_{k}> 0$ irrespective of \(f\), and also local configurations satisfying \(\abs{\ell} < \abs{\ell}\) do not dissipate. Taking into account that some intermediate fluxes $\Phi_\ell$ are non-monotonic in terms of \(f\), one could expect that they would present a less significant contribution. Inspired by evidence from simulations, we observe the predominance of \(\Phi_{\ell^*}\) and \(\Phi_{k}\), hence we introduce the contribution of these two fluxes as $\sigma_{\ell^*,k}=-\beta_{\ell^*}\Phi_{\ell^*}-\beta_k\Phi_{k}>0$. This represents an approximation but not a bound since the remaining fluxes can change their signs.

\begin{figure}
\hspace{-0.6cm}
\includegraphics[width=\columnwidth]{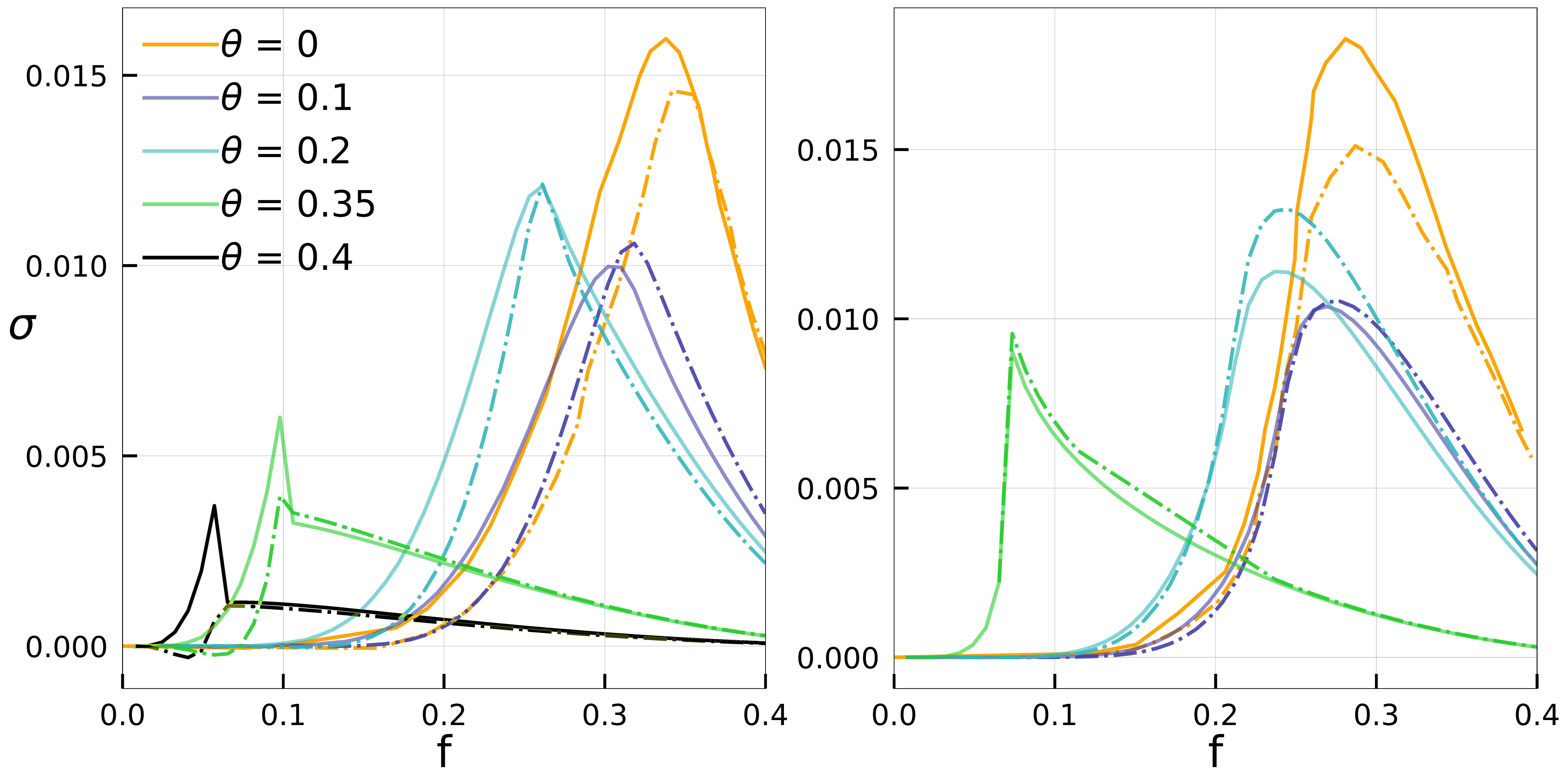} 
\caption{For $k=20$, random-regular (left) and regular (right) structures of sizes $N=1600$ and $40^2$, curves for $\sigma_{\ell^*,k}$ (dot-dashed) and $\sigma$ (continuous) are shown in terms of \(f\) for distinct $\theta$'s. From top to bottom, ${\ell^*}=2,4,6,12$ and $14$.}
\label{fig5}
\end{figure}

Figure~\ref{fig5} compares, for the random-regular and regular lattices, $\sigma_{\ell^*,k}$ and $\sigma$ for distinct values of $\theta$. In all cases, $\sigma_{\ell^*,k}$ is not only close to $\sigma$ but also captures the qualitative behavior, successfully describing the interplay between the control parameter \(f\), inertia \(\theta\), and the dissipation, including a peak located at the vicinity of the phase transition. For larger \(\theta\) the set of dissipating local configurations shrinks, hence the better agreement between both curves.

\section{Conclusion}
The nonequilibrium thermodynamic theory of the generic majority vote model was presented and thoroughly investigated, encompassing its phase transition. A consistent definition of temperature and the connection between heat fluxes and entropy production were introduced and analyzed in the context of continuous and discontinuous phase transitions. The present approach for fluxes is thermodynamically consistent and equivalent to the microscopic entropy production definition and satisfies the detailed fluctuation theorem.

We believe that the present framework not only conciliates the thermodynamic aspects of an important class of nonequilibrium systems but also introduces a new kind of nonequilibrium ingredient, based on the idea of a thermal bath associated with the system neighborhood. Such an idea has revealed general for a generic voter-like model with ``up-down" $Z_2$ symmetry. In the presence of inertia, the spin changes induced by some local configurations are reversible. Moreover, we explore what are the most relevant neighborhoods driving the system dissipation, including its qualitative features across a phase transition, and how these neighborhoods contribute to the structure of the phase diagram.

Our findings are valid for a class that describes systems from social dynamics to the physics of thermal engines, presenting collective effects that can be leveraged for improved performance.
Such potential application raises interesting questions such as the role of lattice topology and even the kind of voter model used (see e.g. Ref. \cite{tome2022stochastic} for a comparison between them) in order to optimize the desirable power and efficiency. Such topics should be investigated in the future.

\bibliography{refs,refs2}
\end{document}